\documentclass[]{spie}  

\usepackage{amsmath,amsfonts,amssymb}
\usepackage{graphicx}
\usepackage[colorlinks=true, allcolors=blue]{hyperref}
\usepackage{algorithm}
\usepackage{algpseudocode}
\usepackage[table]{xcolor}
\usepackage{lastpage}

\title{Distributed learning for automatic modulation recognition in bandwidth-limited networks}

\author[a]{Narges Rashvand}
\author[b]{Kenneth Witham}
\author[a]{Gabriel Maldonado}
\author[a]{Vinit Katariya}
\author[c]{Aly Sultan}
\author[c]{Gunar Schirner}
\author[a]{Hamed Tabkhi}
\affil[a]{Dept. of Electrical and Computer Engineering, University of North Carolina at Charlotte, USA}
\affil[b]{Kostas Research Institute at Northeastern University, Boston, USA}
\affil[c]{Dept. of Electrical and Computer Engineering Northeastern University, Boston, USA}

\authorinfo{Further author information: (Send correspondence to Narges Rashvand)\\E-mail: nrashvan@uncc.edu }

\begin{document} 
\maketitle

\begin{abstract}
Automatic Modulation Recognition (AMR) is critical in identifying various modulation types in wireless communication systems. Recent advancements in deep learning have facilitated the integration of algorithms into AMR techniques. However, this integration typically follows a centralized approach that necessitates collecting and processing all training data on high-powered computing devices, which may prove impractical for bandwidth-limited wireless networks.

In response to this challenge, this study introduces two methods for distributed learning-based AMR on the collaboration of multiple receivers to perform AMR tasks. The TeMuRAMRD 2023 dataset is employed to support this investigation, uniquely suited for multi-receiver AMR tasks. Within this distributed sensing environment, multiple receivers collaborate in identifying modulation types from the same RF signal, each possessing a partial perspective of the overall environment.

Experimental results demonstrate that the centralized-based AMR, with six receivers, attains an impressive accuracy rate of 91\%, while individual receivers exhibit a notably lower accuracy, at around 41\%. Nonetheless, the two proposed decentralized learning-based AMR methods exhibit noteworthy enhancements. Based on consensus voting among six receivers, the initial method achieves a marginally lower accuracy. It achieves this while substantially reducing the bandwidth demands to a 1/256th of the centralized model. With the second distributed method, each receiver shares its feature map, subsequently aggregated by a central node. This approach also accompanies a substantial bandwidth reduction of 1/8 compared to the centralized approach. These findings highlight the capacity of distributed AMR to significantly enhance accuracy while effectively addressing the constraints of bandwidth-limited wireless networks.

\end{abstract}


\section{INTRODUCTION}
\label{sec:intro}  
In wireless communication systems, a variety of modulation types can be employed by the transmitter to manage both data rate and bandwidth utilization. While the transmitter selects the modulation type, the receiver may or may not possess this information. Consequently, modulation details can be embedded within each signal frame, enabling the receiver to recognize the modulation type and respond accordingly. However, due to the limited frequency spectrum, this approach may not be sufficiently efficient in practical scenarios, as it affects spectrum efficiency by introducing additional information into each signal frame\cite{jdid2021machine}. Therefore, AMR as a crucial bridge between signal detection and demodulation, fulfills an essential role in the detection of modulation schemes and eliminating potential overhead in the network protocol. The importance of AMR has been recognized in a wide variety of applications, such as signal demodulation, spectrum sensing, signal surveillance, interference localization, and cognitive radio.\cite {liu2020intelligent, zhang2021efficient}.

In wireless networks, the rapid variations in signal propagation and radio environments persist even when multiple receivers observe the same signal from a single transmitter. Throughout the transmission process, signals emitted by transmitters frequently undergo alterations within the radio frequency channel, such as noise, multi-path fading, shadow fading, center frequency offset, and sample rate offset \cite{zhang2022deep}. So, even for the same signal, receivers that are distributed in a geographical environment have a partial and noisy view of the transmitted signal. In such an environment, achieving efficient modulation recognition requires the transmission of complete observation signals from the receivers to the central node. However, the constraints of bandwidth in bandwidth-limited networks make this impractical \cite{liu2020intelligent}.

Techniques for AMR can be categorized into two distinct main groups: the likelihood-based (LB) method and the feature-based (FB) method. The LB method has the potential to produce great results, but it relies on having prior knowledge of the probability density function (PDF), which introduces significant complexities. In contrast, the FB methods, primarily concentrate on extracting relevant features without the need for additional channel or signal information and involve simpler computational tasks, heavily relying on the process of feature extraction\cite{jdid2021machine, wang2021generalized}.

Over the past few years, deep learning (DL) has achieved significant progress, demonstrating remarkable capabilities in diverse domains from smart cities \cite {10293461,icores24} to computer vision \cite {gholami2023federated1, gholami2023federated2} and signal processing.\cite{wang2021generalized,moukafih2020neural}.
Furthermore, it has been increasingly applied in wireless communication applications, particularly in DL-based AMR. DL-based AMR methods can be seen as feature-based approaches, achieving this by directly extracting signal features through its neural network structure. Nevertheless, the majority of DL-based algorithms for AMR, depend on extensive data stored on a central node, which is gathered from various local clients (receivers), referred to as centralized AMR (CentAMR) \cite{ke2021real, zhang2021efficient, rashvand2024enhancing}. 
As CentAMR relies on a single central node to handle a large amount of data from individual receivers, it faces challenges in increased communication overhead during data transmission and requires a substantial volume of data and considerable computational resources for training \cite{liu2020intelligent}. 
 
In contrast to CentAMR in this domain, which adopts a centralized approach and involves training on signals from diverse receivers, local AMR (LAMR) is defined as training exclusively on signals from a single receiver. However, this localized approach often shows much lower classification accuracy and robustness due to the constraints of limited data and finite computational capabilities.

Due to the limitations imposed by decentralized data storage, available bandwidth, and computing capabilities, CentAMR and LAMR may not be suitable for distributed receivers operating within bandwidth-limited wireless networks. This is where the efficacy of distributed learning-based AMR (DAMR) approaches becomes apparent. These approaches involve the decision-making process while maintaining datasets distributed across local receivers, rather than centralizing them within a single central node. With this approach, local receivers collaboratively contribute to the identification of the modulation type. When DAMR is compared to CentAMR and LAMR, it presents some notable advantages: it eliminates the need for data centralization, decreases demands on bandwidth and computational resources, and enhances the accuracy of local receivers by collaborating with other distributed receivers. Additionally, DAMR can utilize the combined computational capabilities of multiple receivers, reducing the computational burden on a central node in CentAMR. These advantages align well with the inherent features of distributed receivers in bandwidth-limited wireless networks. 


However, most of the studies on DL-AMR and DAMR approaches mainly focus on systems with a single transmitter and receiver pair, which differs from the conditions encountered in real-world applications, where scenarios involve multiple distributed receivers, each with its distinct receive channel effects like multi-path fading and noise.
Recognizing this gap, we have chosen to employ the TeMuRAMR2023 dataset for our research. To tackle AMR in bandwidth-limited networks, we propose two distributed approaches. These approaches are based on consensus voting and feature sharing, utilizing Convolutional Neural Network (CNN) structures. The primary goal is to reduce bandwidth requirements while maintaining a level of accuracy comparable to that of a centralized model.



\section{Related Works}
Radio encoding classification presents a complex problem due to the variability in features. Many vastly different encoding protocols exist \cite{Oetting1979}, some of which have varying levels of complexity (ex. 16QAM, 32QAM, 64QAM, 128QAM, 256QAM). Before deep learning, pattern recognition \cite{Gardner1980} and feature extraction \cite{Gardner1981} algorithms were designed by hand for each modulation type. This design methodology is time-intensive and not easily extensible as drastically different modulation techniques may not be detectable by the designed classifier's method.  Deep learning has been proposed as a solution to the limitations of previous methods. \cite{osheaConvolutionalRadioModulation2016, Oshea2017, osheaAirDeepLearning2017, tekbiyikRobustFastAutomatic2020} propose methods of directly classifying modulation type as well as methods for generating data sets. \cite{karraModulationRecognitionUsing2017} proposes a hierarchy of classifiers to first classify modulation type and then order. \cite{soltaniSpectrumAwarenessEdge2019} proposes a hierarchy of classifiers that will classify both type and order at higher SNR levels and classify more general types at lower SNR values. These methods all borrow directly from computer vision-targeted classification techniques. Recurrent neural networks have also been explored, known for their capacity to capture temporal dependencies in sequential data \cite{zhangEfficientDeepLearning2021, keRealTimeRadioModulation2021, osheaAirDeepLearning2017, liaoSequentialConvolutionalRecurrent2021}.

However, the previously mentioned DL-based AMR methods are CentAMR methods and most of them are designed primarily for systems with single transmitter/receiver pairs. There is limited research that explores the application of DL methods for AMR with multiple receivers, and even less exploration of distributed approaches in this domain. Taking into account the limitations associated with CentAMR, such as its demanding training requirements, significant communication overhead, necessitating a large dataset, and considerable computational power, deploying CentAMR in the context of wireless networks featuring multiple receivers with restricted bandwidth and scenarios involving the distribution of data among multiple receivers is impractical\cite{wang2020distributed}. A few studies in this domain have concentrated on the implementation of federated learning and distributed learning. Nevertheless, in the majority of these studies, the primary reason for employing distributed learning is privacy protection. As these studies point out, CentAMR poses a significant privacy risk due to the extensive data transfer to the central server for model training. \cite{wang2021federated, wang2021generalized,liu2020intelligent, shi2021signal}. However, it is worth noting that this challenge may not be as significant in wireless networks where signals are accessible. A study by Shi et al.\cite{shi2021signal} explores federated learning-based AMR with differential privacy, specifically concentrating on data privacy in AMR tasks. They show that their achieved accuracy is acceptable compared to the centralized model, while ensuring data security. In a study by Wang et al.\cite{wang2020distributed}, a distributed learning-based approach was introduced as a solution to prevent large-scale leakage of stored data in the server. This study demonstrates that each edge device independently trains its local models using its dataset and shares these local models, instead of sharing their data. The results show that their approach has higher training efficiency compared to CentAMR, and it achieves a similar convergence speed to CentAMR. The importance of different data distributions among nodes, including class imbalance distribution and data mismatch, is explored in other studies by Wang\cite{wang2021federated, wang2021generalized}. Their approach involves the use of distributed learning for AMR with a specific focus on varying data distribution among nodes. Their study shows the classification accuracy comparable to CentAMR while minimizing the risk of data leakage on the server\cite{wang2021federated, wang2021generalized}.

In the earlier mentioned studies, the main reason for adopting a distributed learning approach is privacy concerns, and datasets characterized by a SISO nature are also employed. Yet, in the domain of distributed learning for modulation classification, a dataset is essential where each receiver receives the signal with unique channel effects compared to other receivers in the wireless network. Subsequently, the following section describes the AMR with multiple receivers and provides insights into a dataset having features that realistically represent the characteristics of AMR in this domain.

\section{Problem description and dataset}
\subsection{AMR in Single Transmitter, Multiple Receivers Domain}

\begin{figure}[htp]
\centering
\vspace{-5pt}
\includegraphics[trim=20pt 10pt 10pt 10pt, clip, width=5in]{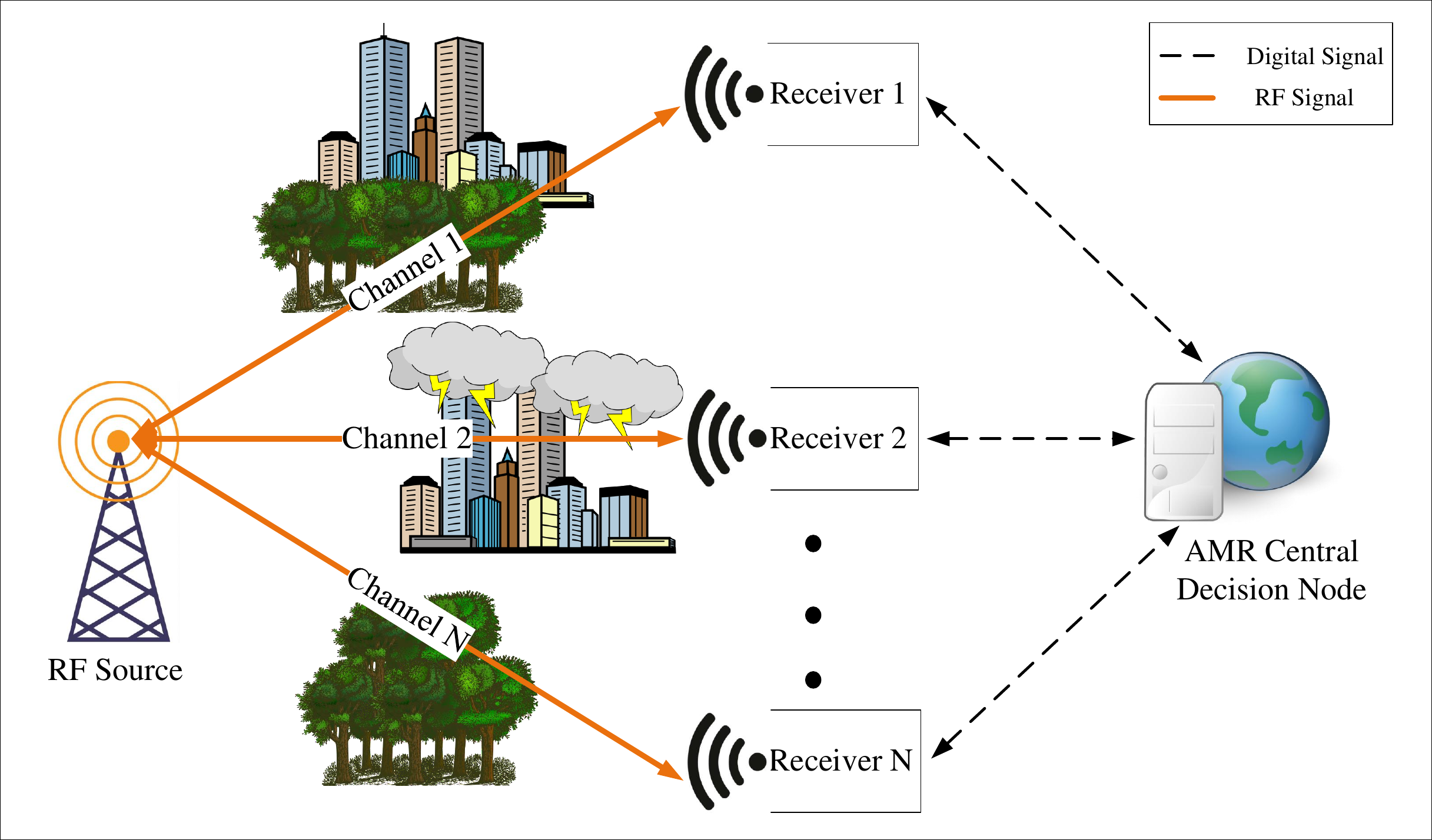}
\caption{AMR with a single transmitter and multiple receivers. Each receiver receives a noisy and partial view of the transmitted signal, enabling a collaborative approach to perform the AMR task effectively.}
\vspace{-5pt}
\label{fig_receivers}
\end{figure}

Figure \ref{fig_receivers} illustrates the schematic of a Single Transmitter Multiple Receivers AMR system. In this configuration, multiple static receivers observe the same transmitted signal from a radio frequency (RF) transmission source. Each of these receivers receives the signal through its own channel, resulting in a partial and noisy view of the transmitted signal. The goal is to collaboratively perform the AMR task, with the objective of identifying the modulation scheme employed by the RF source using a complex IQ time-domain vector. 

As AMR refers to the classification task of various modulation types from received signals, it can be defined by using the Maximum A Posterior (MAP) algorithm. This method can be expressed mathematically as \ref{eq_0} where $\hat{m}_{j}$ represents the predicted modulation type for the received signal, $y_{j}$,  at the jth receiver. The modulation candidate pool is represented by $M$, comprising the modulation types in the dataset. Our modulation pool includes eight modulation types introduced in the dataset section. The mapping function, $F_{DL}$, links the In-phase Quadrature (IQ) samples of the received signal at the jth receiver to the predicted labels, $\hat{m}_{j}$. In short, this formula describes the selection of the predicted modulation type by choosing the one with the highest probability, determined by the DL model.

\begin{equation}
 \hat{m}_{j} = argmax_{n\in M} {F_{DL} (m|y_{j})}
 \label{eq_0}
\end{equation}

\subsection{Dataset }

TeMuRAMRD.2023 (Terrain-driven Multi-Receiver Automatic Modulation Recognition  Dataset) has been presented as a realistic multi-receiver dataset. This dataset has been designed as the Terrain-driven Multi-Receiver Automatic Modulation Recognition Dataset, addressing the need for a realistic multi-receiver dataset. It plays a crucial role in studies involving multi-point radio signals, providing a multi-receiver view of the same signal transmitted by a single transmitter over realistically modeled propagation channels. It contains 8 different digital modulation modes, totaling 896,000 signal samples. Each modulation type has signals with a length of 1024 I/Q samples. The 3GPP 5G channel model \cite{3gpp.38.901} (as implemented by NVIDIA's Sionna \cite{hoydisSionnaOpenSourceLibrary2023} software library) is utilized to simulate propagation channels including multipath and Doppler components. The multi-receiver components of the 3GPP model are also used to simulate cross-correlation effects in the environment and assign each channel its unique channel effect filter. 
This dataset is crucial to our research, which focuses on distributed learning in the presence of multiple receivers. The parameters of this dataset are described in Table \ref{tab_dataset}.

\begin{table}[htpb]
\centering
\caption{Description of TeMuRAMRD.2023 parameters}
\begin{tabular}{|c|p{6cm}|} 
\hline
\textbf{Number of Modulation Types} &  8 \\ \hline
\textbf{Modulation Pool} & BPSK, QPSK, 8PSK, DQPSK, MSK, 16-QAM, 64-QAM, 256-QAM \\ \hline
\textbf{Sample Length} & 1024\\ \hline
\textbf{SNR Range} &  -44 dB to +50 dB\\ \hline
\textbf{Number of Samples} &  896,000\\ \hline
\textbf{Format of Each Data Points} &  IQ samples\\ \hline
\textbf{Maximum Number of Receivers} & 6\\ \hline
\textbf{3GPP 5G Channel Models} & Urban Macro and Rural \\ \hline
\end{tabular}
\label{tab_dataset}
\end{table}

\section{Preliminaries}
Before exploring the proposed DAMR methods, it is essential to provide an insight into certain aspects. First, we offer an overview of the neural network model utilized in all experiments. Additionally, we provide an overview of the LAMR and CentAMR methods, using them as a basis for comparison with our proposed DAMR approaches. 

\textbf {A) {VTCNN2-1D Model:}}
In this paper, we utilize a convolutional neural network model named VTCNN2-1D, which is a modified structure of the model presented in \cite{hauser2017signal} . The VTCNN2-1D model consists of two one-dimensional convolutional layers followed by two fully connected layers. The structure of VTCNN2-1D is shown in Table \ref{tab_1}, where "Conv1D" is a typical 1-dimensional convolutional layer, and "FCN" represents a fully connected layer. The initial layer consists of 256 channels and a filter size of 7, followed by a second layer with 80 channels and a filter size of 7. The model ends with two dense layers, one with 256 neurons and the other with 8-class neurons. The structure of this model is presented in Table \ref{tab_1}.

\begin{table}[htpb]
\centering
\caption{The structure of VTCNN2-1D, including four layers: two convolutional layers and two fully connected layers}
\begin{tabular}{|cc}
\hline
\multicolumn{1}{c|}{\textbf{No.}} & \textbf{Structure} \\ \hline \hline
\multicolumn{1}{c|}{1} & Conv1D(256,7)\\ \hline
\multicolumn{1}{c|}{2} & Conv1D(80,7)\\ \hline
\multicolumn{1}{c|}{3} & FCN(256)\\ \hline
\multicolumn{1}{c|}{4} & FCN (8)\\ \hline
\end{tabular}
\label{tab_1}
\end{table}

\textbf {B) {LAMR method:}}
LAMR methods rely on the receivers, where each receiver is more than just a passive receiver and it possesses a degree of intelligence. In this method, each receiver trains its model based on its dataset, and the modulation recognition is performed locally at each receiver. As illustrated in Figure \ref{fig_receivers}, upon receiving signals from the RF source, each individual receiver independently analyzes its received signal to determine the modulation type, with no information exchange among receivers. Consequently, this method experiences minimal communication overhead. However, since each receiver predicts the modulation type based on its partial view of the signal, it might not achieve the best recognition accuracy. This limitation arises from each receiver's lack of access to diverse signal observations from different perspectives, coupled with computational power constraints preventing the utilization of complex models.

\textbf {C) {CentAMR method:}}
CentAMR methods are generally trained on the central node using large datasets from all local receivers. While this approach can lead to optimal classification results, it also results in significant communication costs during the inference phase, due to the fact that local receivers are not equipped with any form of intelligence and are unable to carry out the AMR task independently. Therefore, the received signal needs to be transmitted to the central node before demodulation. Moreover, the central node faces a computational load due to handling extensive data.

\begin{figure}[htp]
\centering
\vspace{-5pt}
\includegraphics[angle=-90, trim=30pt 30pt 90pt 20pt, clip, width=5.0in]{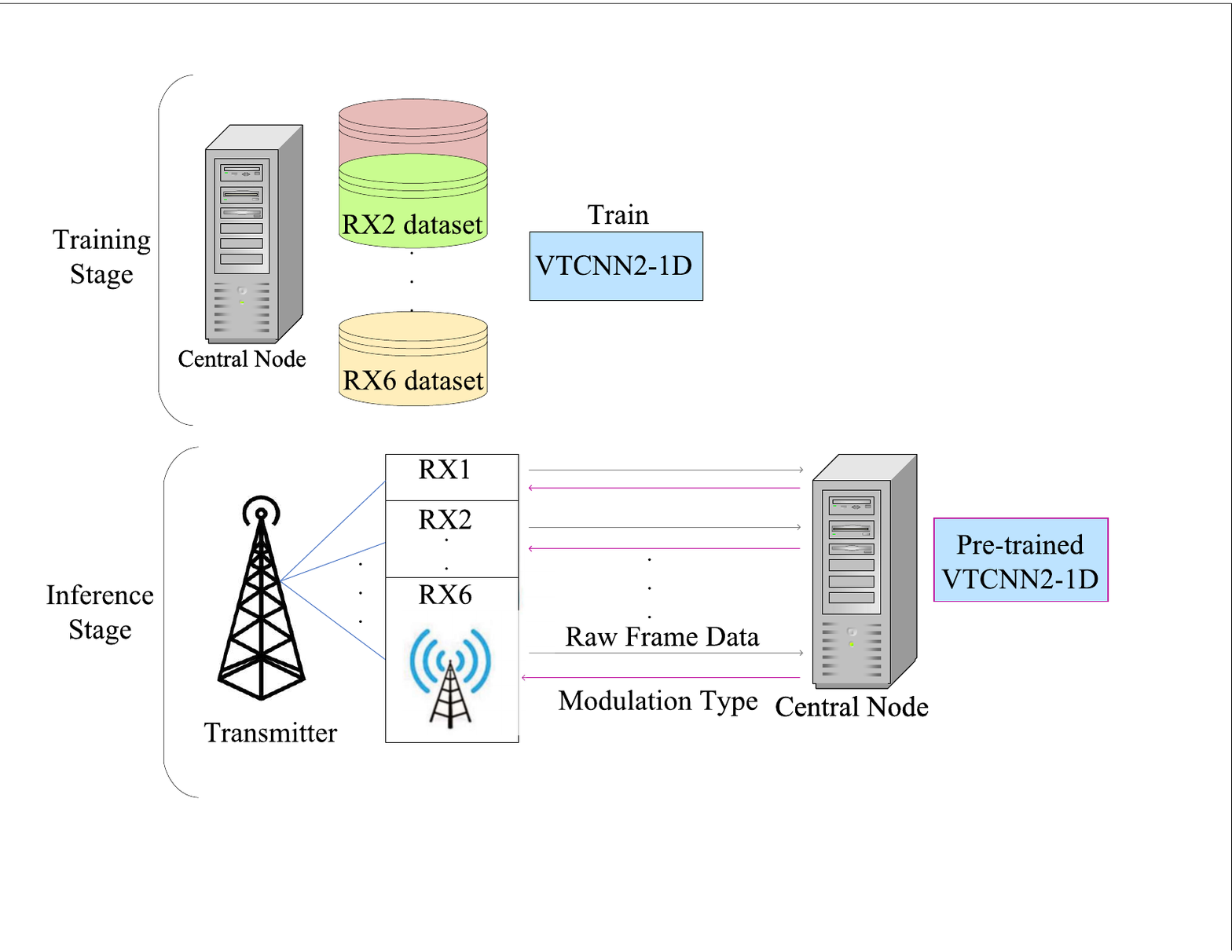}
\caption{
A visual illustration of the CentAMR approach, highlighting how each receiver, referred to as RX, sends the IQ samples to the central node. This central node employs a pre-trained model, previously trained on data collected from all six receivers, to recognize the modulation type used in the received signals. Then, the predicted modulation type is sent to the receivers.}
\vspace{-5pt}
\label{fig_DL1}
\end{figure}

Figure \ref{fig_DL1} illustrates the CentAMR involving a single transmitter and a set of six local receivers. In this configuration, only one VTCNN2-1D is trained using the combined datasets from different local receivers, as opposed to LAMR with training multiple VTCNN2-1D models for each corresponding receiver.
So, initially, the VTCNN2-1D undergoes training on six integrated receiver datasets. The central node employs this trained model to make predictions for modulation classification.
During the inference stage, the transmitter broadcasts the same signal to all six receivers, each experiencing its distinct channel distortion. Upon reception, each receiver converts the signal to a digitalized format and transmits its own raw frame data to the central node. This data includes 1024 samples for the in-phase (I) and quadrature (Q) components. The central node receives six pairs of I and Q samples, each pair representing diverse observations of the same signal from different perspectives obtained by six receivers. Subsequently, the central node utilizes the pre-trained VTCNN2-1D to predict the modulation type of this signal. In the final step, the modulation type is transmitted to all the receivers, enabling them to demodulate the signal.

For the CentAMR method, we employ the same CNN structure based on VTCNN2-1D, as described in Table \ref{tab_1} and we implemented structural adjustments in the initial layer of this model. The VTCNN2-1D has two input channels designed for a single receiver. Given the presence of six receivers, the centralized model's input channels need to be modified to a total of 12.

\section{Proposed distributed AMR Approaches}
\subsection{Distributed AMR (DAMR) methods }

DAMR allows the identification of signal modulation types when datasets are distributed across local receivers instead of gathering them into a single central node. In this approach, each receiver possesses a degree of intelligence and is required to transmit specific information to the central node, avoiding the need to send the entire frame data in the form of IQ samples. This section provides an explanation of our two proposed DAMR methods. 

\subsubsection{Consensus voting method (DAMR-V)}

In this section, we introduce one of the cooperative decision rules based on consensus voting, which is a subset of ensemble methods in machine learning. "Ensembling" is an area focused on increasing model capacity without significantly raising computational demands. The ensemble methods are utilized to combine the outputs of various machine learning models based on specific rules. Different decision fusion strategies can be employed to combine these ensemble models, and the choice depends on how they combine the outputs of the ensemble models. These strategies include methods like majority voting, model averaging, stacked generalization, and more\cite {ganaie2022ensemble,mahabub2019voting,moukafih2020neural}.

\begin{figure}[htp]
\centering
\includegraphics[angle=-90, trim=230pt 50pt 200pt 350pt, clip, width=5.5in]{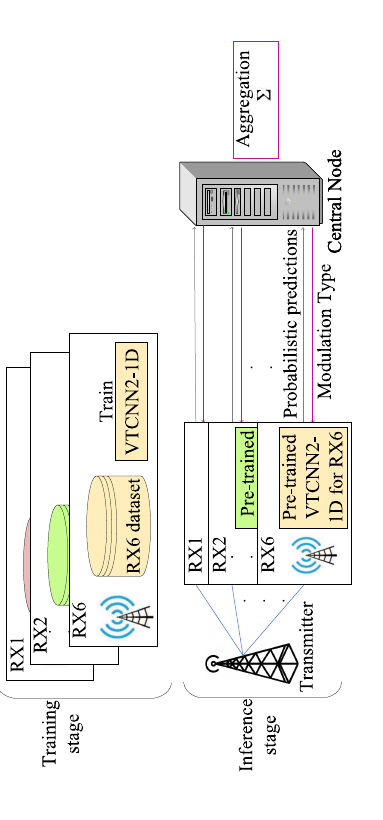}
\caption{The DAMR-V method, where each receiver trains a local model and shares their probabilistic predictions. The central node then combines these predictions to determine the modulation type, which is subsequently sent back to the receivers.}
\vspace{-10pt}
\label{fig_DL2}
\end{figure}

The structure of the first proposed method, DAMR-V, is depicted in Figure \ref{fig_DL2}. This figure illustrates how receivers collaboratively identify the modulation type of the same received signal without the need to exchange their I and Q samples with the central node.

During the training stage of this method, each receiver independently trains their local VTCNN2-1D models using their corresponding datasets. Since VTCNN2-1D is trained for a single receiver setup, it only has two input channels.
In the inference stage, each individual receiver, upon receiving the signal and converting it to I and Q samples, employs its pre-trained model, which was previously trained on its datasets. The I and Q samples are then fed into their pre-trained model and the probabilistic predictions for modulation types are transmitted to the central node.
The central node, acting as the decision maker, collects and combines these probabilistic predictions obtained from all receivers and computes their summation, as described in Equation \ref {eq:1}, where $N_{rx}$ denotes the total number of receivers, $W_j$ represents the weight that can be assigned to the $j$th receiver. It should be highlighted that as each receiver is fair, and the decision at each receiver is equally important, thus no specific weights are assigned to individual receivers.

\begin{equation}
 \hat{P} = \sum_{j=1}^{N_{rx}} W_j P_{j} \label{eq:1}
\end{equation}

This cumulative probability, denoted as $\hat{P}$, becomes the basis for the identification of the modulation type by the central node. Referring to equation \ref{equ1}, the final decision, $\hat{y}$, is determined by selecting the modulation type with the highest combined probability obtained from the set of modulation types, $M$. Subsequently, each receiver retrieves the predicted modulation type for signal demodulation.

\begin{equation}
\hat{y}=argmax_{n\in M} \hat{P}
\label{equ1}
\end{equation}


\subsubsection{Feature Sharing based method (DAMR-F)}

\begin{figure}[htp]
\centering
\vspace{-5pt}
\includegraphics[trim=60pt 230pt 230pt 250pt, clip, width=10.5in]{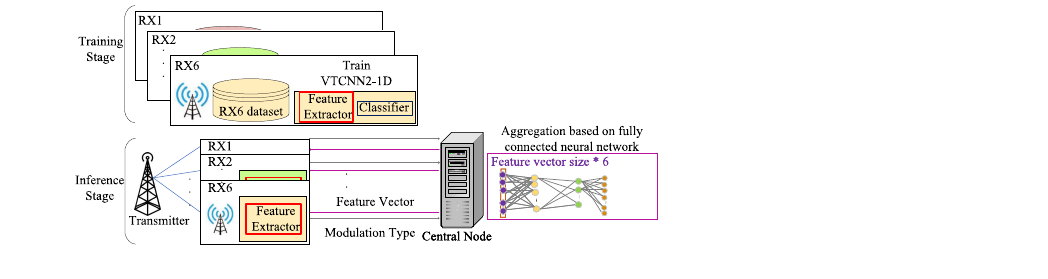}
\caption{Illustration of DAMR-F, where individual receivers train local models, exchange features, and the central node aggregates these features for prediction via fully connected layers. The predicted modulation types are then sent back to the receivers for signal demodulation.}
\vspace{-10pt}
\label{fig_DL3}
\end{figure}

In this section, an alternative distributed learning approach for AMR is introduced, based on feature sharing. The structure of this method is illustrated in Figure \ref{fig_DL3}. This process involves the receivers collaboratively determining the modulation type of the received signal by exchanging the features they have extracted.

In the training stage of this method, similar to the first approach, each receiver independently trains their local VTCNN2-1D model using their corresponding datasets. This trained model is subsequently used during the inference stage.
In the inference stage, receivers extract features from the received signal using their individual trained model and share these features with the central node. The central node aggregates these collected features, resulting in 
a feature set that $N_{rx}$ times larger than the feature set held by any single receiver. Utilizing a model composed of fully connected layers, the central node recognizes the modulation type of the signal. Upon determining it, this information is sent back to all the receivers so they can demodulate the signal properly.

\section{Simulation Results and Discussions}
\subsection{Classification Accuracy and Bandwidth Usage}
\label{s6.1}
In our experiments, the TeMuRAMRD.2023 dataset was partitioned as follows: 60\% for training, 20\% for testing, and the remaining 20\% allocated to validation. The framework for all the experiments was PyTorch, utilizing a Cross Entropy (CE) loss function which is the common loss function for classification tasks. In addition, the batch size is set as 256. The simulations are performed on a Tesla V100 GPU.
We selected two AMR methods for comparison: CentAMR and LAMR, as described in the preliminaries section.

Table \ref {tab_2} presents the classification accuracy results of the LAMR method, where each receiver trains the model exclusively on their local datasets independently. As observed from Table \ref {tab_2}, the accuracy achieved using this approach varies between 37.54\% and 43.48\%. Thus, the LAMR demonstrates poor performance, with an average classification accuracy of 41.28\% across the six receivers. This limitation arises from each receiver having only a partial view of the transmitted signal.

\begin{table}[htpb]
\centering
\caption{Classification Accuracy for LAMR for six receivers}
\begin{tabular}{|cc}
\hline
\multicolumn{1}{c|}{\textbf{Receivers}} & \textbf{Classification Accuracy} \\ \hline \hline
\multicolumn{1}{c|}{First Receiver} & 42.60\%\\ \hline
\multicolumn{1}{c|}{Second Receiver} & 37.54\%\\ \hline
\multicolumn{1}{c|}{Third Receiver} & 43.48\%\\ \hline
\multicolumn{1}{c|}{Fourth Receiver} & 42.37\%\\ \hline
\multicolumn{1}{c|}{Fifth Receiver} & 41.47\% \\ \hline
\multicolumn{1}{c|}{Sixth Receiver} & 40.26\% \\ \hline
\end{tabular}
\label{tab_2}
\end{table}

Furthermore, Table \ref{tab_3} illustrates the average classification performance for CentAMR, DAMR-V, DAMRF-256, and DARMF-8. The CentAMR model is trained on a collection of six sub-datasets collected from the receivers. CentAMR demonstrates impressive performance with an average classification accuracy of up to 91\%. This represents a significant 50\% improvement from LAMR to CentAMR. This enhancement is attributed to the processing of the received signal from different perspectives provided by each receiver. In the case of the DAMR-F method, we conducted experiments with varying numbers of features acquired from individual receivers. Following the architecture of VTCNN2-1D, after extracting features through convolutional layers, the trained model of each receiver produces 8 and 256 features, referred to as DAMR-F8 and DAMR-F256, respectively. For the DAMR-F8, the central node aggregates 8 features from each of the six receivers, resulting in a total of 48 features. The fully connected neural network in this configuration comprises an initial layer with 48 neurons and two hidden layers, with 128 and 64 neurons, respectively. Additionally, a dropout of 0.3 is applied after the last hidden layer. For DAMR-F256, the central node gathers 256 features from every receiver, totaling 1536 features. We implement a more complex model compared to the fully connected layers used for DAMR-F8, consisting of three hidden layers with 4096, 1024, and 512 neurons, respectively. Dropout is also applied after the second and third hidden layers.

As evident in Table \ref {tab_3}, the proposed DAMR methods showcase the ability to enhance classification performance without necessitating data sharing among local receivers. Both DAMR methods demonstrate notable efficacy, achieving an accuracy of around 89\%. DAMR-V achieves this accuracy with each receiver exclusively sharing the probabilistic predictions, while DAMR-F8 and DAMR-F256 attain the almost same accuracy by sharing 8 and 256 features among local receivers and the central node. DAMR methods have a huge breakthrough in the classification performances, when compared with the LAMR, while DAMR still has a slight performance gap with the CentAMR. The performance gap between the CentAMR and the DAMR methods is just about 2\%. The accuracy for DAMR-F8 and DAMR-F256 is almost identical and the increasing number of features shared between the local receivers and the central node does not contribute to a significantly improved accuracy. This lack of accuracy improvement is due to the absence of additional information from the increased features. It suggests the possibility that a more sophisticated model may be required to achieve further enhancements in performance.

\begin{table}[htpb]
\centering
\caption{Classification Accuracy and Bandwidth Usage: A Comparative Analysis during Inference Stage for CentAMR, DAMR-V, DAMR-F8 and DAMR-F256}
\begin{tabular}{|c|c|c}
\hline
\multicolumn{1}{c|}{\textbf{Methods}} & \textbf{Classification Accuracy} & \textbf{Bandwidth Usage}  \\ \hline \hline
\multicolumn{1}{c|}{CentAMR} & 91.07\% & 64 kbits\\ \hline
\multicolumn{1}{c|}{DAMR-V} & 89.24\% (\textcolor{blue}{1.83\%$\downarrow$})&  256 bits  (\textcolor{red}{99.6\%$\downarrow$})  \\ \hline
\multicolumn{1}{c|}{DAMR-F8} & 89.20\% (\textcolor{blue}{1.87\%$\downarrow$})& 256 bits   (\textcolor{red} 
  {99.6\%$\downarrow$}) \\ \hline
\multicolumn{1}{c|}{DAMR-F256} & 89.32\% (\textcolor{blue}{1.75\%$\downarrow$})& 8 kbits  (\textcolor{red}{87.5\%$\downarrow$}) \\ \hline
\end{tabular}
\label{tab_3}
\vspace{3pt} 
\begin{tabular}{@{}p{\linewidth}@{}}
        Note: The blue font in the table indicates the decrease in accuracy for the proposed DAMR methods and the red font corresponds to the bandwidth reduction achieved with the proposed DAMR in comparison to CentAMR.
    \end{tabular}
\end{table}

As observed in Table \ref {tab_3}, the bandwidth requirement for each modulation classification in CentAMR is 64 Kbits. To elaborate on the calculation of this, each receiver needs to transmit raw frame data to the central node. Each frame comprises 1024 IQ samples in a  32-bit floating-point format, resulting in a total bandwidth requirement of 64 Kbits for identifying the modulation type from the received signal.
In DAMR-V, every local receiver transmits its probability prediction to the central node in the format of a 32-bit floating point number. Considering the existence of 8 modulation types, this transmission necessitates a total of 256 bits. In DAMR-F8, every receiver shares 8 features with the central node, summing up to 256 bits. As for DAMR-F256, where each receiver shares 256 features, the requirement for each modulation identification is 8 Kbits. Therefore, the bandwidth savings achieved by DAMR-V and DAMR-F8 are  $\frac{1}{256}$th of CentAMR, while DAMR-F256 attains a $\frac{1}{8}$th reduction compared to CentAMR.

\begin{figure}[htp]
\centering
\vspace{0pt}
\includegraphics[trim=20pt 30pt 20pt 20pt, clip, width=5in]{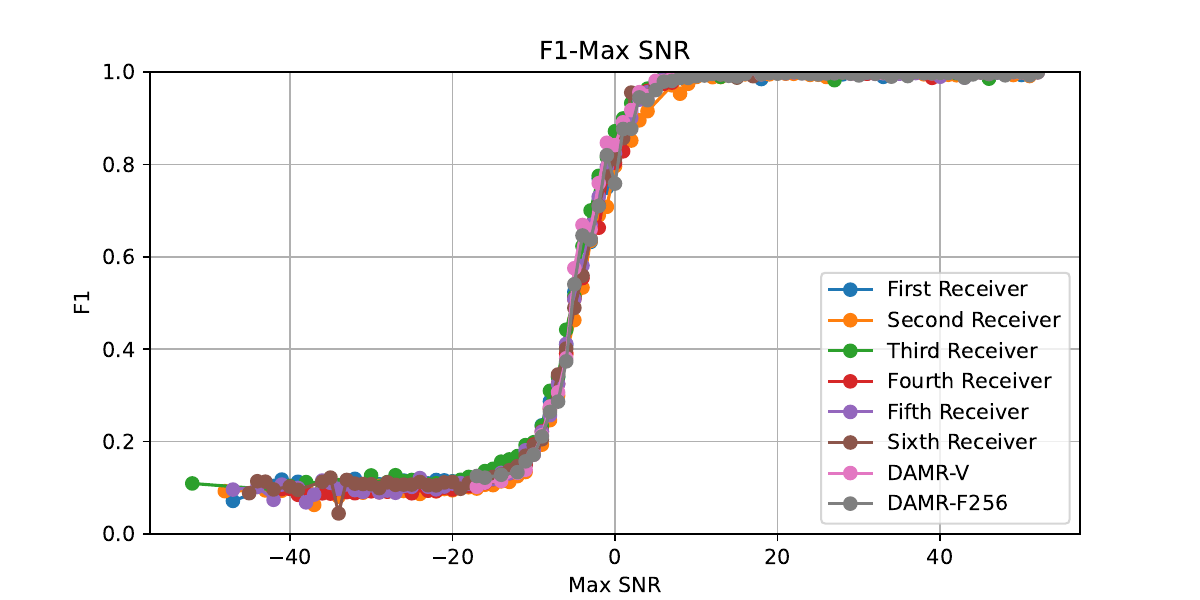}
\caption{F1 versus SNR for LAMR, DAMR-V, DAMR-F256.}
\vspace{0pt}
\label{fig_F1_SNR}
\end{figure}

The curves displaying classification performances versus SNR are illustrated in Figure \ref{fig_F1_SNR}. In creating this figure for DAMR, we selected the highest SNR across all receivers. While DAMR methods led to a notable improvement in average accuracy, as indicated in Table \ref{tab_3}, these methods did not have accuracy enhancement across varying SNR levels, as can be seen in Figure \ref{fig_F1_SNR}. Considering this, one possible hypothesis is that VTCNN2-1D model's simplicity could be inadequate. In response to this observation, we broadened our experiments to include a scenario involving a more complex DL model for AMR task. The objective is to investigate potential improvement in accuracy under varying SNR conditions by increasing the number of receivers.

\begin{figure}[htp]
\centering
\vspace{+5pt}
\includegraphics[trim=20pt 0pt 20pt 10pt, clip, width=5in]{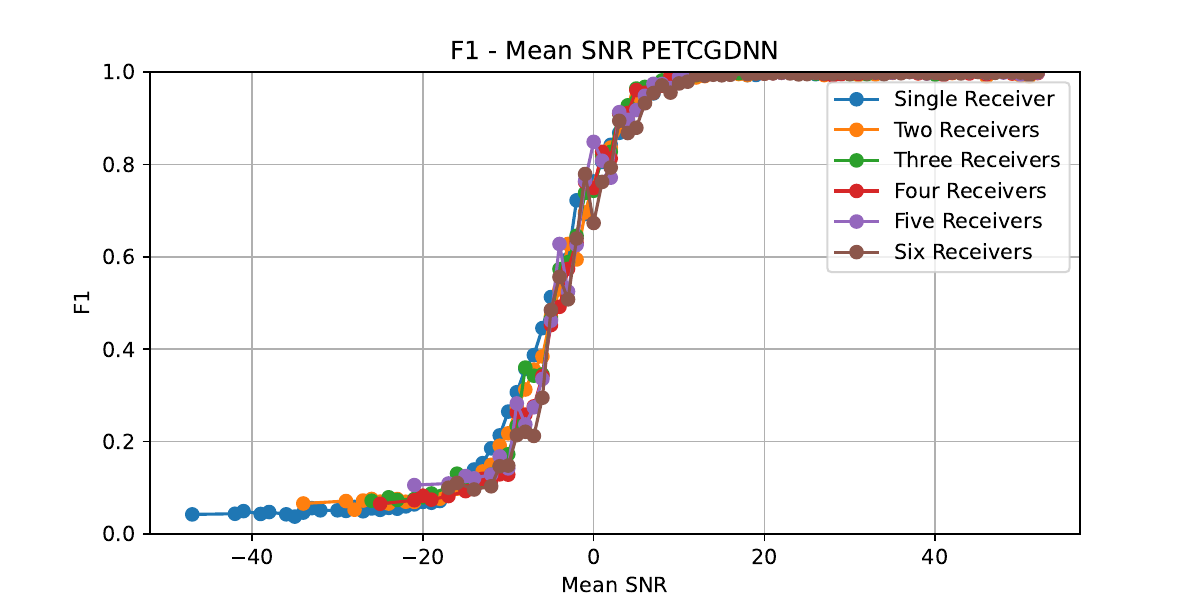}
\caption{Comparing F1 versus SNR for different number of receivers based on PET-CGDNN, ranging from 1 to 6.}
\vspace{-10pt}
\label{fig_F1_SNR_PETCGDNN}
\end{figure}

Our selected model to explore this hypothesis is the PET-CGDNN model which is presented in \cite {zhang2021efficient}. This model is based on phase parameter estimation and transformation, incorporating convolutional neural networks (CNN) and gated recurrent units (GRA) as feature extraction layers. Initially, we evaluate the model's performance on TeMuRAMRD.2023, where it demonstrates higher average accuracy compared to the model used in our experiments, VTCNN2-1D. Subsequently, we assess its performance with an increasing number of receivers, ranging from 1 to 6. The F1 versus SNR is illustrated in  Figure \ref{fig_F1_SNR_PETCGDNN}, highlighting the ongoing challenge of achieving higher accuracy at the same SNR level, even with a mode complex model.

\subsection{Latency and Throughput}

In our comprehensive analysis, we compare the performance of the CentAMR against DAMR-V and DAMR-F8, focusing on the important metrics of latency and throughput. Latency is defined generally as the time required for a model to process an individual input, with the measurement expressed in seconds. This metric is essential for understanding a model's responsiveness and immediacy in executing tasks. Throughput, on the other hand, measures the model's ability to process a specific number of inputs within a designated period. We conducted our latency and throughput experiments using the NVIDIA Jetson AGX Orin Developer Kit. It has the GPU with 2048 CUDA cores, a 12-core ARM-based CPU, and 64 GB of memory. The NVIDIA JetPack version utilized was 6.0-b52. To carry out our measurements, we utilized the CPU with the MAXN performance mode of the kit.

The results of our detailed assessments on latency and throughput of the CentAMR, DAMR-V, and DAMR-F8 are presented in Table \ref{tab_4}. CentAMR has a latency of  17.46 ms/sample, whereas DAMR-V and DAMR-F8 show latency of 76.06 and 92.44 ms/sample, respectively. However, CentAMR achieves its lower latency, leading to better throughput, at the expense of significant communication overhead. This overhead results in bandwidth usage that is 256 times greater than 
DAMR-V and 8 times higher than DAMR-F8, as discussed in Section \ref{s6.1}. The slightly lower latency of DAMR-V (76.06 ms/sample) compared to latency of DAMR-F8 (92.44 ms/sample) can be attributed to the simpler aggregation process in the central node. Specifically, DAMR-V employs addition, whereas DAMR-F8 utilizes a fully connected neural network for aggregation.

\begin{table}[htpb]
\centering
\caption{Measured latency and throughput of CentAMR, DAMR-V, and DAMR-F8  on NVIDIA Jetson AGX Orin Developer Kit.}

\begin{tabular}{|c|c|c}
\hline
\multicolumn{1}{c|}{\textbf{Methods}} & \textbf{Latency (ms/sample)} & \textbf{Throughput (sample/second)}  \\ \hline \hline
\multicolumn{1}{c|}{CentAMR} & 17.46 &  57.27 \\ \hline
\multicolumn{1}{c|}{DAMR-V} & 76.06  & 13.15 \\ \hline
\multicolumn{1}{c|}{DAMR-F8} & 92.44 & 10.82\\ \hline

\end{tabular}
\label{tab_4}
\vspace{3pt} 
\end{table}

\section{Conclusion}
In this paper, we address the challenges of AMR in wireless networks with the limitations on bandwidth. We explore scenarios involving multiple receivers observing the same signal from a single transmitter, facing different channel effects and leading to partial and noisy signal views. The TeMuRAMRD.2023 was utilized to train models on multi-point radio signals. 
Methods like LAMR and CentAMR have limitations, prompting the introduction of two distributed AMR approaches. Experimental results highlight the significant advancements brought by DARM methods, highlighting its potential in real-world applications, particularly in bandwidth-limited wireless networks. The proposed DAMR methods, DAMR-V and DAMR-F, eliminate the need for data centralization, reduce demands on bandwidth and computational resources, and enhance accuracy through collaboration among local receivers. The simulation results indicate a minor performance difference between DAMR and CentAMR. These DAMR methods leverage distributed learning, achieving an impressive average classification accuracy of around 89\%, just 2\% below CentAMR, while reducing the bandwidth requirement to 1/256th and 1/8th of the CentAMR. 


In our future research, we aim to explore federated learning-based approaches. This method involves individual participants sharing their model weights, showcasing significant potential, especially in bandwidth-limited wireless networks. Additionally, we plan to investigate the significance of transformer models, which have demonstrated their efficacy in handling sequential data. Our upcoming research also involves an exploration of distributed learning grounded in transformers. This next step aims to uncover the potential of transformers in extracting more meaningful information from received signals for each receiver and to understand how this shared information can be combined to enhance overall accuracy.


\bibliography{report} 
\bibliographystyle{spiebib} 

\end{document}